\documentclass[a4paper,fleqn,usenatbib]{mnras}

\usepackage{newtxtext,newtxmath}
\usepackage[T1]{fontenc}
\usepackage{ae,aecompl}

\usepackage{graphicx}	% Including figure files
\usepackage{amsmath}	% Advanced maths commands
\usepackage{amssymb}	% Extra maths symbols
\usepackage{siunitx}    % numbers/unit formatting
\usepackage{bbm} 		% More maths symbols
\usepackage[dvipsnames]{xcolor}		% Text colors and the like
\usepackage{soul} 		% For colorboxes/highlighting of text
\usepackage{booktabs}	% Better tables
\usepackage{array}		% More tables options
\usepackage[inline]{enumitem}	% Options for lists
\usepackage{multirow}   % Yet more tables options
\usepackage{etoolbox}   % Needed for below command that patch style for author/year citing to only colour/link year (from http://hosting.astro.cornell.edu/~tleung/wordpress/?p=287)
\sethlcolor{yellow}

\newcommand{\xp}{X_{\textrm{p}}} % stress prior to glitch X_p
\newcommand{\rhot}{\rho_{\Delta t}} % autocorrelation between consecutive waiting times
\newcommand{\rhox}{\rho_{\Delta X}} % autocorrelation between consecutive sizes
\newcommand{\rhob}{\rho_{\,\textrm{b}}} %cross-correlation between size and backward waiting time
\newcommand{\rhof}{\rho_{\,\textrm{f}}} %cross-correlation between size and forward waiting time

\makeatletter
\patchcmd{\NAT@citex}
  {\@citea\NAT@hyper@{%
     \NAT@nmfmt{\NAT@nm}%
     \hyper@natlinkbreak{\NAT@aysep\NAT@spacechar}{\@citeb\@extra@b@citeb}%
     \NAT@date}}
  {\@citea\NAT@nmfmt{\NAT@nm}%
   \NAT@aysep\NAT@spacechar\NAT@hyper@{\NAT@date}}{}{}
\patchcmd{\NAT@citex}
  {\@citea\NAT@hyper@{%
     \NAT@nmfmt{\NAT@nm}%
     \hyper@natlinkbreak{\NAT@spacechar\NAT@@open\if*#1*\else#1\NAT@spacechar\fi}%
       {\@citeb\@extra@b@citeb}%
     \NAT@date}}
  {\@citea\NAT@nmfmt{\NAT@nm}%
   \NAT@spacechar\NAT@@open\if*#1*\else#1\NAT@spacechar\fi\NAT@hyper@{\NAT@date}}
  {}{}
\makeatother

\title[Glitch size and waiting time autocorrelations]{Autocorrelations in pulsar glitch waiting times and sizes}

\author[Carlin et al.]{
J. B. Carlin,$^{1}$
A. Melatos$^{1,\,2}$
\\
$^{1}$School of Physics, University of Melbourne, Parkville, VIC 3010, Australia\\
$^{2}$Australian Research Council Centre of Excellence for Gravitational Wave Discovery (OzGrav), \\ \ University of Melbourne, Parkville, VIC 3010, Australia\\
}

\date{Accepted XXX. Received YYY; in original form ZZZ}

\pubyear{2019}

\begin{document}
\label{firstpage}
\pagerange{\pageref{firstpage}--\pageref{lastpage}}
\maketitle

\begin{abstract}
Among the five pulsars with the most recorded rotational glitches, only PSR J0534$+$2200 is found to have an autocorrelation between consecutive glitch sizes which differs significantly from zero (Spearman correlation coefficient $\rho=-0.46$, p-value $=0.046$). No statistically compelling autocorrelations between consecutive waiting times are found. The autocorrelation observations are interpreted within the framework of a predictive meta-model describing stress-release in terms of a state-dependent Poisson process. Specific combinations of size and waiting time autocorrelations are identified, alongside combinations of cross-correlations and size and waiting time distributions, that are allowed or excluded within the meta-model. For example, future observations of any ``quasiperiodic'' glitching pulsar, such as PSR J0537$-$6910, should not reveal a positive waiting time autocorrelation. The implications for microphysical models of the stress-release process driving pulsar glitches are discussed briefly.
\end{abstract}


%%%%%%%%%%%%%%%%%%%%%%%%%%%%%%%%%%%%%%%%%%%%%%%%%%

%%%%%%%%%%%%%%%%% BODY OF PAPER %%%%%%%%%%%%%%%%%%

\section{Introduction}
The secular electromagnetic spin down of some rotation-powered pulsars is interrupted stochastically by spin-up events called ``glitches''. The statistical properties of glitches have been studied across the whole pulsar population (see \citealp{Shemar1996, Lyne2000, Fuentes2017}; among others), and more recently in individual pulsars as the number of recorded glitches has grown \citep{Melatos2008, Espinoza2011, Ashton2017, Howitt2018, Melatos2018}; see Table \ref{tab:ac} for a list of the main objects studied by previous authors. The latter analyses reveal that there are two main statistical classes of glitching pulsar:  ``Poisson-like'' objects, with exponentially distributed waiting times and power-law distributed sizes; and ``quasi-periodic'' objects, which have non-monotonic waiting time and size distributions \citep{Melatos2008, Espinoza2011, Howitt2018}. What physically triggers glitches, and why two classes of activity exist, are open questions. In general terms, most models posit that glitches occur when the elastic stress and/or differential rotation in the star exceed a threshold, triggering some sort of scale-invariant avalanche process such as a starquake or superfluid vortex avalanche; see the recent review by \citet{Haskell2015} and references therein. 

Pulsar glitches are events that are naturally ordered in time. It is therefore profitable to ask whether their order of occurrence contains statistical information about the underlying physics. The ordered set of glitch epochs shows some evidence of clustering, or equivalently a variable rate, in PSR J0534$+$2200 (also known as B0531$+$21) \citep{Lyne2015, Carlin2019a}. Analysis of time-ordered stochastic events is a rich field of study. For example, Omori's law describes the observed sequence of aftershocks following a large terrestrial earthquake \citep{Utsu1995}. Autocorrelations between waiting times of stochastic events have been studied in the context of numerical sandpile simulations \citep{DeMenech2000, Santra2007}, solar flares \citep{Paczuski2005}, and other self-organized critical systems \citep{Caruso2007}. 

\citet{Melatos2018} studied the forward and backward cross-correlations between glitch sizes and waiting times in the context of a state-dependent Poisson process \citep{Daly2007, Wheatland2008, Fulgenzi2017} and made falsifiable predictions regarding the cross-correlation coefficients as functions of the spin-down rate and mean waiting time. In this paper we ask whether falsifiable predictions can also be made regarding sizes and waiting time autocorrelations. In Section \ref{sec:obs} we outline the current observational situation on this front and calculate autocorrelation coefficients for the five pulsars with the most recorded glitches. Section \ref{sec:sec3} sets up the state-dependent Poisson process model for glitches and predicts the autocorrelation coefficient as a function of key inputs to the model, e.g. the spin-down rate. In Section \ref{sec:discussion} we directly compare the theory and existing observations and make falsifiable predictions regarding future observations.

\section{Timing observations}
\label{sec:obs}
\subsection{Data}
Large-scale, multi-object radio timing campaigns devoted to systematic searches for pulsar glitches are currently carried out at the Jodrell Bank \citep{Espinoza2011} and Parkes \citep{Yu2013, Yu2017} Observatories. These campaigns are supplemented by additional current programs such as CHIME \citep{Ng2018} and UTMOST \citep{Jankowski2019} that take place at the Dominion Radio Astrophysical Observatory and Molonglo Synthesis Telescope respectively. The analysis in this paper combines the above observations with historical data sets from the Hartebeesthoek Radio Astronomy Observatory \citep{Buchner2008}, Mount Pleasant Radio Observatory \citep{Palfreyman2016}, Arecibo Observatory \citep{Arzoumanian2018}, and Jet Propulsion Laboratory \citep{Downs1981}. The completeness of the Parkes data set, i.e. whether all detectable glitches have been identified, was discussed by \citet{Yu2017}. \citet{Espinoza2014} claimed that the data set for PSR J0534$+$2200 is complete, and that the minimum physically allowed glitch size is resolved. However, for most pulsars the cadence of observations is variable \citep{Janssen2006, Yu2017}. It is still uncertain whether the data sets we analyze in this paper are complete. 

According to the Jodrell Bank online catalogue\footnote{Found through the Jodrell Bank Centre for Astrophysics at \url{http://www.jb.man.ac.uk/pulsar/glitches.html} \citep{Espinoza2011}.}, the five most prolific glitchers as of 2019 February 11 are PSR J0537$-$6910 ($N=42$ recorded glitches\footnote[2]{The number and parameters of glitches recorded for this pulsar vary between \citet{Middleditch2006}, \citet{Ferdman2017}, and \citet{Antonopoulou2018}. We opt to include in our analysis events that occur in two out of three sources.}), PSR J1740$-$3015 ($N=36$), PSR J0534$+$2200 ($N=23$ or 27\footnote[3]{\label{foot:f3}The first four glitches in the Jodrell Bank catalogue occurred before high-cadence monitoring of PSR J0534$+$2200 began, and there is a known gap in observations between the fourth and fifth recorded glitches \citep{Lyne2015}. Henceforth we denote the full data set with an asterisk (i.e. PSR J0534$+$2200*), and the 23 events since 1982 without an asterisk.}), PSR J1341$-$6220 ($N=23$), and PSR J0835$-$4510 ($N=20$). The mean number of glitches per year are 3.2, 1.1, 0.64, 1.1, and 0.38 for the five objects respectively; they have been monitored for different lengths of time.

\subsection{Autocorrelations}
We can arrange the epochs, $t_i$, and fractional sizes, $s_i = \Delta \nu_i / \nu_i$, in any given pulsar as a sequence of time-ordered waiting times between glitches, \{$\Delta t_1$, $\Delta t_2$, ..., $\Delta t_{N-1}$\}, with $\Delta t_i = t_{i+1} - t_{i}$, and a sequence of time-ordered sizes, \{$s_1$, $s_2$, ..., $s_{N}$\}. Note that, if $N$ glitches are observed, there are $N-1$ observed waiting times. The autocorrelation of an ordered data set of $N$ discrete points, $\{x_i\}$, is calculated by constructing $N-k$ pairs of points, $(x_1, x_{1+k}), (x_2, x_{2+k}), ..., (x_{N-k}, x_N)$, i.e. pairs of points separated by lag $k$. The first and second entries in each pair constitute the two variates to be correlated. The Spearman rank correlation coefficient, $\rho$, is calculated as \citep{Lehmann2006}
\begin{equation}
\label{eq:spearman}
{1 - \rho = 6\, \bigg\{(N-k)\left[(N-k)^2 - 1\right] \bigg\}^{-1} \sum_{i=1}^{N-k} d_i^2} \ \ ,
\end{equation}
where $d_i$ is the difference between the ordinal ranks of the \emph{i}-th pair of observations. While it is possible to calculate autocorrelations at an arbitrary lag $k$, we restrict our subsequent analysis to $k=1$, i.e. autocorrelations between consecutive events. A partial check of lags $1 < k \leq 5$ does not reveal any autocorrelations significantly different from zero. We use the Spearman rank correlation coefficient, which looks for monotonic relationships, instead of the standard Pearson correlation coefficient, as the former is less sensitive to outliers and does not assume a parametric (e.g. linear) form for the relationship.

\begin{figure*}
	\centering %
	\includegraphics[width=0.9\linewidth]{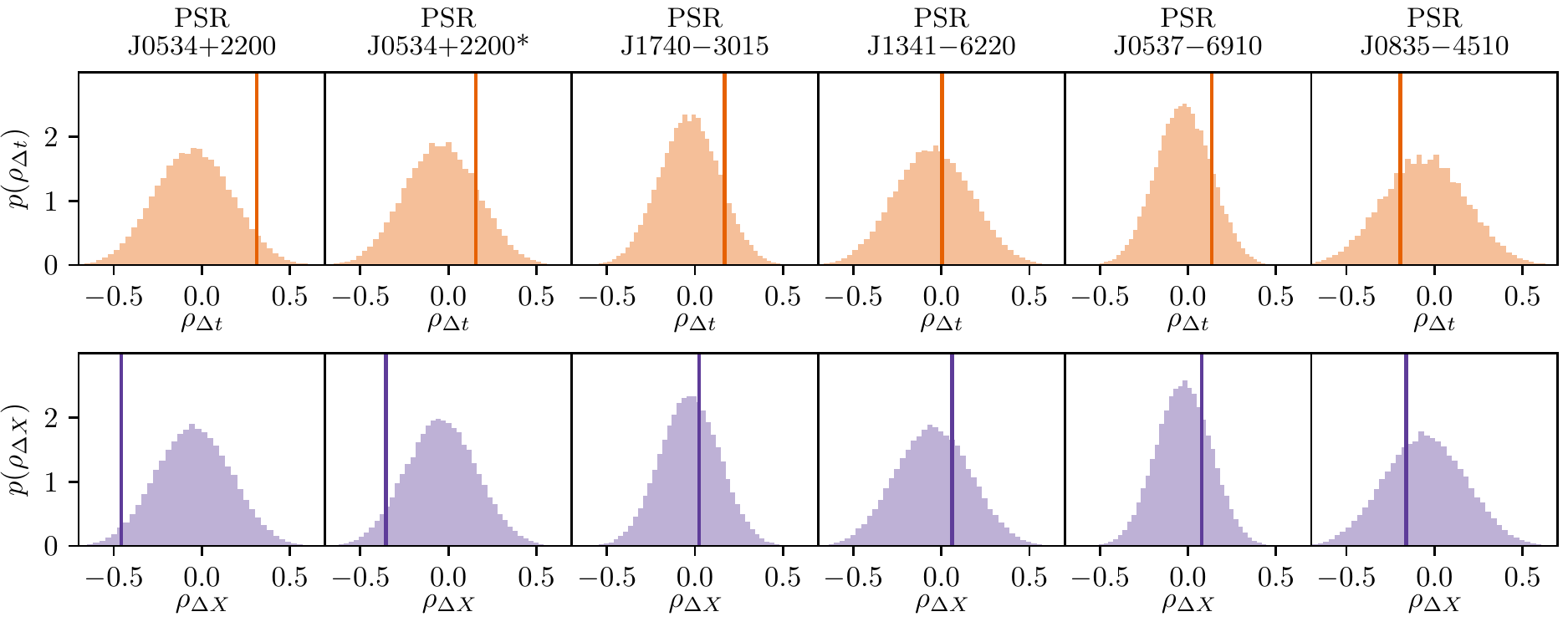}
	\caption{Measured autocorrelation coefficient between consecutive waiting times, $\rhot$ (top row, orange vertical lines), and consecutive sizes, $\rhox$ (bottom row, purple vertical lines), for the five most active glitching pulsars. The null distribution (colored histogram in each panel) is estimated using a bootstrap permutation method, as described in the text. See Table \ref{tab:ac} for numerical values of $\rhot$ and $\rhox$.}
	\label{fig:all5_plot}
\end{figure*}

Table \ref{tab:ac} contains the calculated Spearman correlation coefficients for autocorrelations in consecutive waiting times, $\rhot$, and consecutive sizes, $\rhox$, for the five most active glitching pulsars. The autocorrelations for PSR J0534$+$2200* (see footnote \ref{foot:f3}) are calculated by ignoring the comparison between $s_4$ and $s_5$, and between $\Delta t_3$ and $\Delta t_5$, in \eqref{eq:spearman} due to the known gap in observations between the fourth and fifth glitch \citep{Lyne2015}. The significance of the calculated Spearman correlation coefficients is estimated using a bootstrap permutation method. This nonparametric method uses permutations of the ordered data set to estimate the null distribution (i.e. the distribution of $\rho$, if there is no autocorrelation in the data). We use this estimate of the null distribution to calculate a p-value: the probability that we would see $|\rho|$ greater than the calculated value, if the null hypothesis is true. This method is robust when compared to the asymptotic (i.e. large $N$) or parametric assumptions of other significance tests \citep{Hall1992, Good2006}. None of the six data sets in Table \ref{tab:ac} have autocorrelations in waiting times or sizes that are significantly different from zero (p-value $> 0.05$), barring perhaps the size autocorrelations in PSR J0534$+$2200 ($\rhox=-0.46$, p-value $=0.046$). As we are in effect carrying out 12 independent significance tests it should not be surprising that at least one of the 12 has a p-value of less than $0.05$, if the null hypothesis (that there is no autocorrelation) is true for all data sets. 

Figure \ref{fig:all5_plot} shows the estimated null distributions (shaded histograms) and calculated Spearman correlation coefficients (vertical lines) for the autocorrelation between consecutive waiting times (top row) and sizes (bottom row) in the five most active glitching pulsars. The null distributions are quite broad due to the small number of glitches observed in each pulsar. We do not show the confidence intervals for the measured values of $\rho$ in Figure \ref{fig:all5_plot} for clarity. However they are consistent with the p-values, i.e. the 95\% confidence interval includes the value of $\rho=0$ for all coefficients except the size autocorrelation in PSR J0534$+$2200. 

\begin{table}
	\centering %
	\caption{Pulsar name, number of glitches ($N$), Spearman autocorrelation coefficient ($\rho$), and associated p-value for waiting times (subscript $\Delta t$) and sizes (subscript $\Delta X$) for the five pulsars with the most recorded glitches. The p-value is estimated using a bootstrap permutation method, as described in the text.}
	\begin{tabular}{l c S[table-format=1.2] S[table-format=1.2] S[table-format=1.2] S[table-format=1.2]}
		& & \multicolumn{2}{c}{Waiting times} & \multicolumn{2}{c}{Sizes} \\
	Name (J2000)  & $N$  & {$\rhot$} & {p-value} & {$\rhox$} & {p-value} \\
	\midrule
	PSR J0534$+$2200  & 23 & 0.31   & 0.094 & -0.46 & 0.046 \\
	PSR J0534$+$2200* & 27 & 0.16   & 0.35  & -0.35 & 0.11 \\
	PSR J1740$-$3015  & 36 & 0.17   & 0.25  & 0.024 & 0.75  \\
	PSR J1341$-$6220  & 23 & 0.0026 & 0.81  & 0.059 & 0.62  \\
	PSR J0537$-$6910  & 42 & 0.14   & 0.31  & 0.079 & 0.51  \\
	PSR J0835$-$4510  & 20 & -0.20  & 0.56  & -0.16 & 0.64  \\
	\bottomrule
	\end{tabular}
	\label{tab:ac}
\end{table}	

\section{State-dependent Poisson process}
\label{sec:sec3}
\subsection{Meta-model}
\label{sec:sdpp}
Long-term glitch activity can be meta-modelled as a state-dependent Poisson process without specializing to a particular glitch mechanism \citep{Fulgenzi2017, Melatos2018, Carlin2019b}. The meta-model assumes that the instantaneous glitch rate at time $t$, $\lambda(t)$, is governed by a single variable: the mean-field stress in the star, $X(t)$. The exact nature of $X(t)$ depends on the physical mechanism causing glitches. For example it could be the spatially averaged lag between the angular velocities of the rigid crust and the superfluid interior in the vortex avalanche picture \citep{Anderson1975, Warszawski2011}, or the crustal strain in the starquake picture \citep{Larson2002, Middleditch2006}. It is assumed that $\lambda[X(t)]$ grows monotonically with time, as the stress builds due to spin down, until $\lambda[X(t)]$ diverges at some critical stress $X_\textrm{cr}$, and some fraction of the stress is released. Although we present the meta-model henceforth in terms of the vortex avalanche picture we emphasize that it applies equally to any stick-slip stress-release process \citep{Melatos2018}. 

The equation of motion for the system is
\begin{equation}
{X(t) = X(0) + t - \sum_{i=1}^{N(t)} \Delta X^{(i)}}\ \ ,
\label{eq:eom}
\end{equation}
where $X$ and $t$ are expressed in dimensionless units of $X_{\textrm{cr}}$ and $X_{\textrm{cr}}I_\textrm{c}/N_\textrm{em}$ respectively, $I_\textrm{c}$ is the moment of inertia of the crust, $N_\textrm{em}$ is the electromagnetic torque acting on the crust, and $X(0)$ is an arbitrary initial condition. Both $N(t)$, the number of glitches up to and including time $t$, and the size of each stress-release event, $\Delta X^{(i)}$, are random variables, making the process an example of a doubly stochastic Poisson process \citep{Cox1955, Grandell1976}.

The sizes, $\Delta X^{1}$, \dots, $\Delta X^{N(t)}$, are drawn from a conditional jump distribution, $\eta(\Delta X \mid \xp)$, with $\int d(\Delta X) \eta(\Delta X \mid \xp) = 1$. The function $\eta$ depends explicitly on the stress in the system just prior to the glitch, $\xp$, because we stipulate that no glitch reduces the stress in the system below zero \citep{Fulgenzi2017}. The exact functional form of $\eta$ is unobservable; it depends on the glitch microphysics. \citet{Fulgenzi2017} and \citet{Melatos2018} used a power law with exponent $-1.5$ and fractional lower cutoff $\beta$, but in this paper we allow $\eta$ to vary, following the framework in Section 3 of \citet{Carlin2019b}. 

The counting function $N(t)$ is implicitly determined by repeated draws from the standard probability density function (PDF) for waiting times, $\Delta t$, from a variable rate Poisson process \citep{Cox1955},
\begin{equation}
\label{eq:delt}
{p[\Delta t \mid X(t)] = \lambda\left[ X(t) + \Delta t\right] \exp\left\{ -\int_{t}^{t+\Delta t} \text{d} t' \lambda [X(t')] \right\}}\ \ .
\end{equation}
The exact form of the rate function $\lambda[X(t)]$ does not significantly impact the long-term dynamics of the system \citep{Fulgenzi2017, Carlin2019b}, so long as there is a divergence at the critical lag $X_\textrm{cr}$. Following previous work, we use
\begin{equation}
\label{eq:rate}
{\lambda[X(t)] = \frac{\alpha}{1 - X(t)}}\ \ ,
\end{equation}
where
\begin{equation}
{\alpha = \frac{I_\textrm{c} X_\textrm{cr} \lambda_0}{N_\textrm{em}}}
\end{equation}
is a dimensionless control parameter and $\lambda_0$ is a reference rate, i.e. $\lambda_0 = \lambda(1/2)/2$. 

Long-term glitch statistics are generated by running a Monte-Carlo automaton which alternates drawing $\Delta t$ from \eqref{eq:delt} and $\Delta X$ from $\eta(\Delta X \mid \xp)$ while tracking the stress $X(t)$ and hence $\lambda[X(t)]$. We find that the automaton output falls into two regimes: ``fast'' spin-down ($\alpha \lesssim 1$), which generates power-law distributed sizes and exponentially distributed waiting times, and ``slow'' spin-down ($\alpha \gtrsim 1$), which generates sizes and waiting times distributed with the same functional form as $\eta$ \citep{Fulgenzi2017, Carlin2019b}. 

\citet{Melatos2018} studied the size--waiting-time \emph{cross}-correlations predicted by the above meta-model. When $\eta$ is a power law, the state-dependent Poisson process predicts large positive cross-correlations between sizes and forward waiting times, when $\alpha$ is small, and small positive cross-correlations between sizes and backward waiting times, when $\alpha$ is large; see Figure 4 in \citet{Melatos2018}. When $\eta(\Delta X \mid \xp)$ is not a power law, the large positive cross-correlation between sizes and forward waiting times at small $\alpha$ remains, while the small positive cross-correlation between sizes and backward waiting times at large $\alpha$ increases, depending on which functional form is used, see Appendix A and Table 1 in \citet{Carlin2019b} for details. These falsifiable theoretical trends open the door to a number of interesting observational tests.

\subsection{Autocorrelations: qualitative predictions}
\label{sec:qual}
Does the meta-model outlined in Section \ref{sec:sdpp} predict analogous trends for size and waiting time autocorrelations? We first argue qualitatively that the answer is yes before confirming the result with simulations in Section \ref{sec:quant}. For example, in the fast spin-down regime ($\alpha \lesssim 1$), we have $\eta(\Delta X \mid \xp) \approx \eta(\Delta X \mid 1)$, as the stress in the system quickly recovers to $X \approx 1$ after each glitch. Hence the system does not remember the size of the previous glitch, nor the waiting time between the previous two glitches. There are no size or waiting time autocorrelations in this regime, regardless of the choice of $\eta$. On the other hand, in the slow spin-down regime ($\alpha \gtrsim 1$), we expect different behavior. If $\eta$ is peaked around a fraction, $\mu_\textrm{G}$, of $\xp$, a positive autocorrelation between consecutive sizes (but not waiting times) should arise. In this scenario, $p[\Delta t \mid X(t)]$ does not change much with time, as $X(t)$ remains small, when $\alpha$ is high. Hence, we expect no waiting time autocorrelation, as the waiting times are effectively independent draws from the same PDF. However consecutive glitch sizes are correlated, because a fraction of the current stress is released at each glitch; if the stress is higher than average to begin with, one observes a sequence of larger than average glitches, before the stress resets back to its mean value. 

\subsection{Autocorrelations: quantitative predictions}
\label{sec:quant}
\begin{figure}
	\centering
	\includegraphics[width=\linewidth]{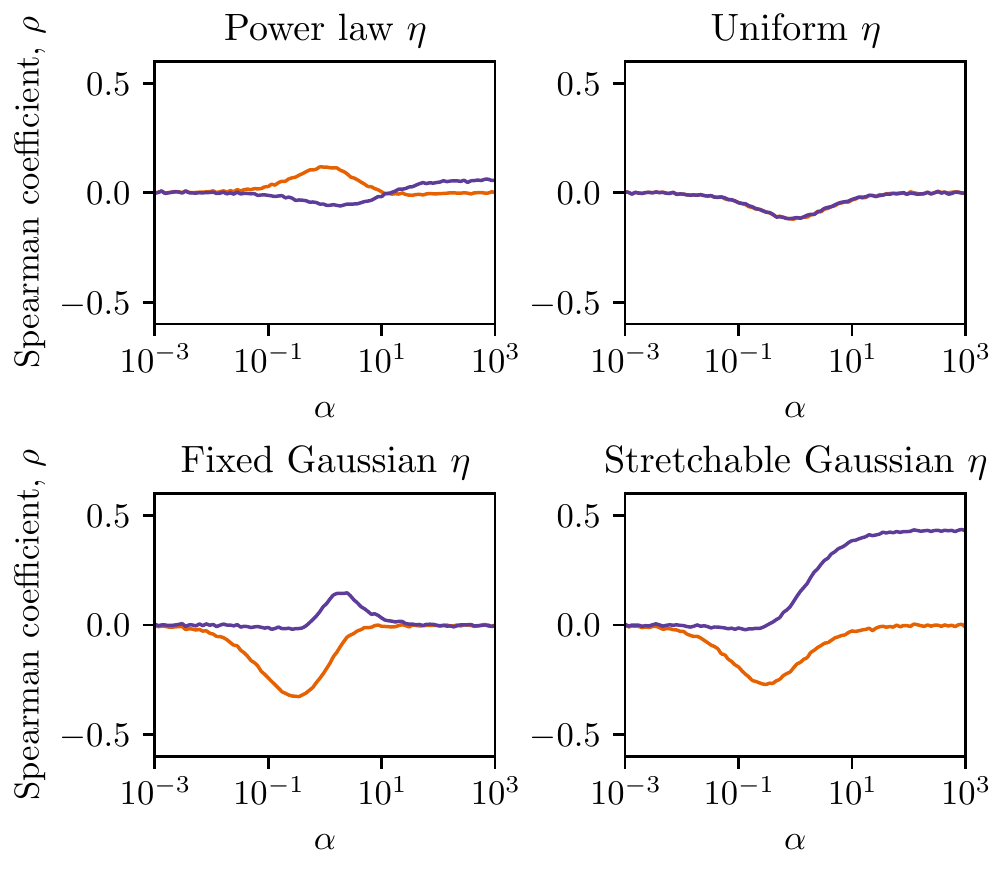}
	\caption{Autocorrelations between consecutive waiting times ($\rhot$, orange curves) and sizes ($\rhox$, purple curves) for glitches generated using the state-dependent Poisson process outlined in Section \ref{sec:sdpp}. Simulation parameters: 100 logarithmically spaced $\alpha$ values, $10^5$ glitches per $\alpha$ value, rate function as given by \eqref{eq:rate}, $\beta=10^{-2}$ for power-law $\eta$ (top-left panel), $\mu_\textrm{G}=0.5$ and $\sigma_\textrm{G}=0.125$ for fixed and stretchable Gaussian $\eta$ (bottom-left and bottom-right panels respectively). The uniform $\eta$ (top right panel) has no free parameters. Explicit functional forms for each $\eta(\Delta X \mid \xp)$ are presented by \citet[Table 1]{Carlin2019b}.}
	\label{fig:acorr_sp}
\end{figure}

To quantify the trends identified in Section \ref{sec:qual}, we run a Monte Carlo automaton to simulate sequences of glitches from the model defined in Section \ref{sec:sdpp}, given $\eta$ and $\alpha$. Pseudocode for the automaton is presented in Section 2.5 in \citet{Carlin2019b}. The functional forms of $\eta$ used in our simulations are the same ones used by \citet{Carlin2019b} to study size--waiting-time cross-correlations. Figure \ref{fig:acorr_sp} shows $\rhot$ (orange curves) and $\rhox$ (purple curves) for four different functional forms of $\eta(\Delta X \mid \xp)$. When $\eta(\Delta X \mid \xp)$ is a power law (top left panel), both $\rhot$ and $\rhox$ are small for all values of $\alpha$. There is a slight rise to $\rhot \approx 0.15$ around $\alpha \approx 1$, which coincides with a slight dip to $\rhox \approx -0.05$. When $\eta(\Delta X \mid \xp)$ is uniform (top right panel), $\rhot \ll 1$ and $\rhox \ll 1$ are identical at all $\alpha$, with a trough of $\rho \approx -0.1$ at $\alpha \approx 1$. When $\eta(\Delta X \mid \xp)$ is Gaussian, the behavior changes. The lower two panels of Figure \ref{fig:acorr_sp} correspond to two types of Gaussian: ``fixed'' (bottom left panel) and ``stretchable'' (bottom right panel). These correspond to a Gaussian that is peaked at a fixed value of $X$, regardless of $\xp$, and a Gaussian that is peaked at a fraction of $\xp$, respectively; see Section 3.1 in \citet{Carlin2019b} for details. For both fixed and stretchable Gaussians we find $\rhot < 0$ for $\num{5e-2} \lesssim \alpha \lesssim 1$. On the other hand, $\rhox$ differs between the two functional forms. When the Gaussian is fixed, $\rhox$ peaks at $\approx 0.15$ at $\alpha \approx 1$, with $\rhox \ll 1$ otherwise. When the Gaussian is stretchable, $\rhox$ grows monotonically with $\alpha$, asymptoting to $\rhox \approx 0.45$ at $\alpha \geq 10^3$.

\begin{figure}
	\centering
	\includegraphics[width=0.7\linewidth]{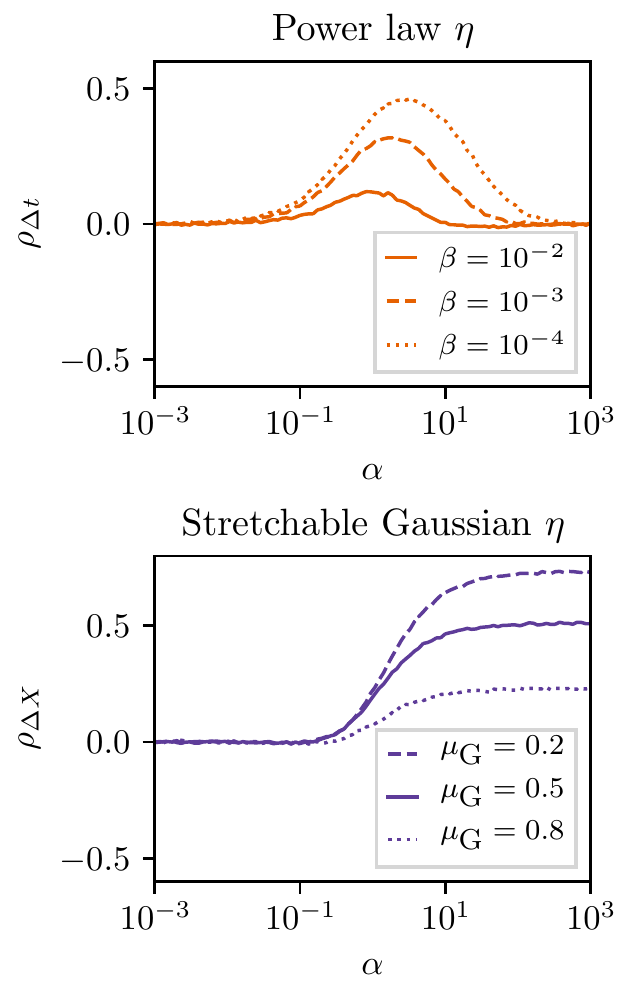}
	\caption{Autocorrelations generated by the state-dependent Poisson process. Top panel: $\rhot$ when $\eta(\Delta X \mid \xp)$ is a power law and $\beta$ is changed from $10^{-2}$ (solid curve) to $10^{-3}$ (dashed curve) to $10^{-4}$ (dotted curve). Bottom panel: $\rhox$ when $\eta(\Delta X \mid \xp)$ is a stretchable Gaussian and $\mu_\textrm{G}$ is changed from 0.2 (dashed curve) to 0.5 (solid curve) to 0.8 (dotted curve). Simulation parameters: 100 logarithmically spaced $\alpha$ values, $10^5$ glitches per $\alpha$ value, rate function as given by \eqref{eq:rate}, power-law index fixed at $-1.5$ (top panel), $\sigma_\textrm{G}=0.125$ (bottom panel). Explicit functional forms for $\eta(\Delta X \mid \xp)$ are presented by \citet[Table 1]{Carlin2019b}.}
	\label{fig:beta_mug}
\end{figure}

When $\eta(\Delta X \mid \xp)$ is a power law, the automaton output depends on $\beta$, the fractional minimum size of a glitch. When $\beta$ is adjusted from $10^{-2}$ to $10^{-3}$ to $10^{-4}$, the peak in $\rhot$ shifts from $\rhot \approx 0.15$ around $\alpha \approx 1$ to $\rhot \approx 0.35$ around $\alpha \approx 2$ to $\rhot \approx 0.50$ around $\alpha \approx 3$, as we see in the top panel of Figure \ref{fig:beta_mug}. On the other hand, $\rhox$ does not change appreciably with $\beta$. The index of the power law also affects $\rhot$. A shallower power law with index $-0.5$ shifts the slight rise at $\alpha \approx 1$ to a slight dip at the same $\alpha$, i.e. the output approaches the case when $\eta$ is uniform. Interestingly, steeper power law indices of $-2$ and $-3$ also produce a small trough in $\rhot$ at $\alpha \approx 0.2$ of $\rhot \approx -0.05$ and $\rhot \approx -0.15$ respectively. 

When $\eta(\Delta X \mid \xp)$ is a Gaussian, the automaton output depends on both the standard deviation of $\eta$, denoted by $\sigma_\textrm{G}$, and the mean, denoted as $\mu_\textrm{G}$. For both the fixed and stretchable Gaussian $\eta$, as $\sigma_\textrm{G}$ increases, the system again approaches the case where $\eta$ is uniform. For the fixed Gaussian the rise in $\rhox$ and the dip in $\rhot$ shifts $\approx0.3$ dex down (up) in $\alpha$ for $\mu_\textrm{G} = 0.8$ (0.2) as compared to $\mu_\textrm{G} = 0.5$. For the stretchable Gaussian, $\mu_\textrm{G}$ is inversely proportional to $|\rhox|$ and $|\rhot|$ at a given $\alpha$. We show the behavior of $\rhox$ with changing $\mu_\textrm{G}$ in the bottom panel of Figure \ref{fig:beta_mug}. With $\mu_\textrm{G} =0.2$ the peak $\rhox$ increases to $\approx 0.7$ and the trough in $\rhot$ decreases to $\approx -0.4$. For $\mu_\textrm{G} = 0.8$ $\rhox$ decreases and $\rhot$ increases. The behavior of $\rhox$ is anticipated qualitatively at high values of $\alpha$, because when $\mu_\textrm{G}$ is low, $X(t)$ spends more time being higher (or lower) than average, at fixed $\alpha$, compared to when $\mu_\textrm{G}$ is high.

\section{What do autocorrelations teach us?}
\label{sec:discussion}
The state-dependent Poisson meta-model is agnostic regarding the exact mechanism underlying glitches. It describes any process that hovers around a point of marginal stability, with events triggered at some threshold, e.g. superfluid vortex avalanches, crustquakes, and many other models commonly proposed in the literature \citep{Haskell2015}. It is therefore profitable to compare its autocorrelation predictions with data, knowing that the conclusions are unlikely to depend on the specific microphysics.
 
\subsection{Existing data}
No significant autocorrelations in sizes or waiting times have been measured to date in the five objects in Section \ref{sec:obs}. What does this tell us about glitch physics when combined with the results in Section \ref{sec:sec3}? It is hard to make firm statements without more data. However the large negative size autocorrelation seen in PSR J0534$+$2200 ($\rhox=-0.46$, p-value $=0.046$) is incongruous with most models in the literature. That is to say, there is no combination of $\alpha$ and $\eta$ that generates $\rhox =-0.46$. Moreover, we see no evidence in any pulsar for a strong, positive size autocorrelation in existing data, as expected if $\eta$ is a stretchable Gaussian and we have $\alpha \gg 1$, (see the bottom-right panel of Figure \ref{fig:acorr_sp}). Finally, existing data disfavor models that predict sizable negative waiting time autocorrelations, as seen if $\eta$ is Gaussian with $0.1 \lesssim \alpha \lesssim 1$.

\subsection{Future data}
As the number of recorded glitches grows, the variance of the null distributions displayed in Figure \ref{fig:all5_plot} shrinks. Armed with accurate measurements of both $\rhot$ and $\rhox$, it will eventually be feasible to rule out sections of the parameter space of the state-dependent Poisson process model for individual pulsars. For example, $\alpha \lesssim 0.1$ is disallowed, if either $\rhot$ or $\rhox$ differ significantly from zero. Similarly, $\alpha \gtrsim 10$ is disallowed, if $\rhot$ is positive. Uniform $\eta$ is ruled out, if either autocorrelation differs significantly from zero, e.g. $|\rho| \geq 0.1$. A positive $\rhot$ is only possible if $\eta$ is a power law. The magnitude of a positive $\rhot$ places constraints on $\alpha$ and $\beta$. Any observed positive $\rhot$ (which implies that $\eta$ is a power law) should come along with a negligible $\rhox$, otherwise the model is not self-consistent.

\subsection{Combining auto- and cross-correlations}
\label{sec:combine}
To test the state-dependent meta-model further, we can con-currently consider size autocorrelations, waiting time autocorrelations, cross-correlations between sizes and backwards waiting times ($\rhob$), and cross-correlations between sizes and forwards waiting times ($\rhof$). \citet{Melatos2018} found that, when $\eta$ is a power law, we should see a large $\rhof$ alongside $\rhob \ll 1$ in the low-$\alpha$ regime, and low $\rhof$ and $\rhob$ in the high-$\alpha$ regime. Similar predictions are made for numerous choices of $\eta$ \citep{Carlin2019b}.

\begin{figure}
	\centering
	\includegraphics[width=0.8\linewidth]{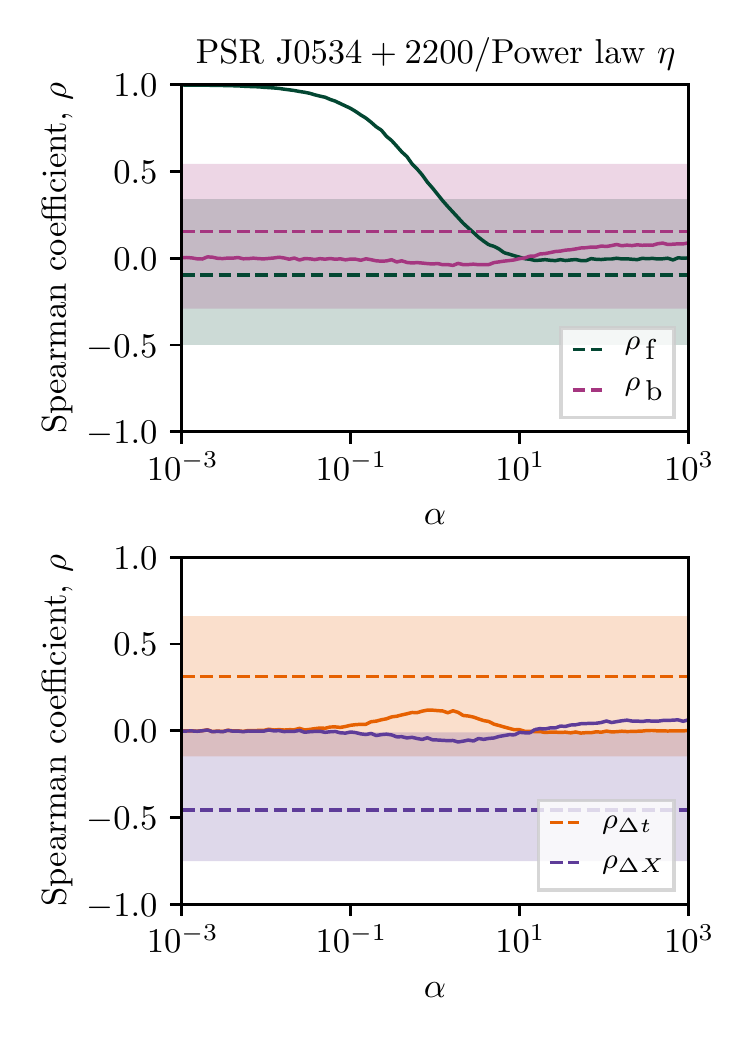}
	\caption{Theoretical (solid curves) and measured (dashed lines and shaded bands) correlations for PSR J0534$+$2200. Top panel: Size--forward waiting time cross-correlation ($\rhof$, green curves), and size--backward waiting time cross-correlation ($\rhob$, pink curves). Bottom panel: Waiting time autocorrelation ($\rhot$, orange curves), and size autocorrelation ($\rhox$, purple curves). The shaded bands indicate the 95\% confidence interval for each observed correlation. Model parameters common to both panels: power law $\eta$, $\beta=10^{-2}$, power-law index of $-1.5$.}
	\label{fig:comp_2200}
\end{figure}

\begin{figure}
	\centering
	\includegraphics[width=0.8\linewidth]{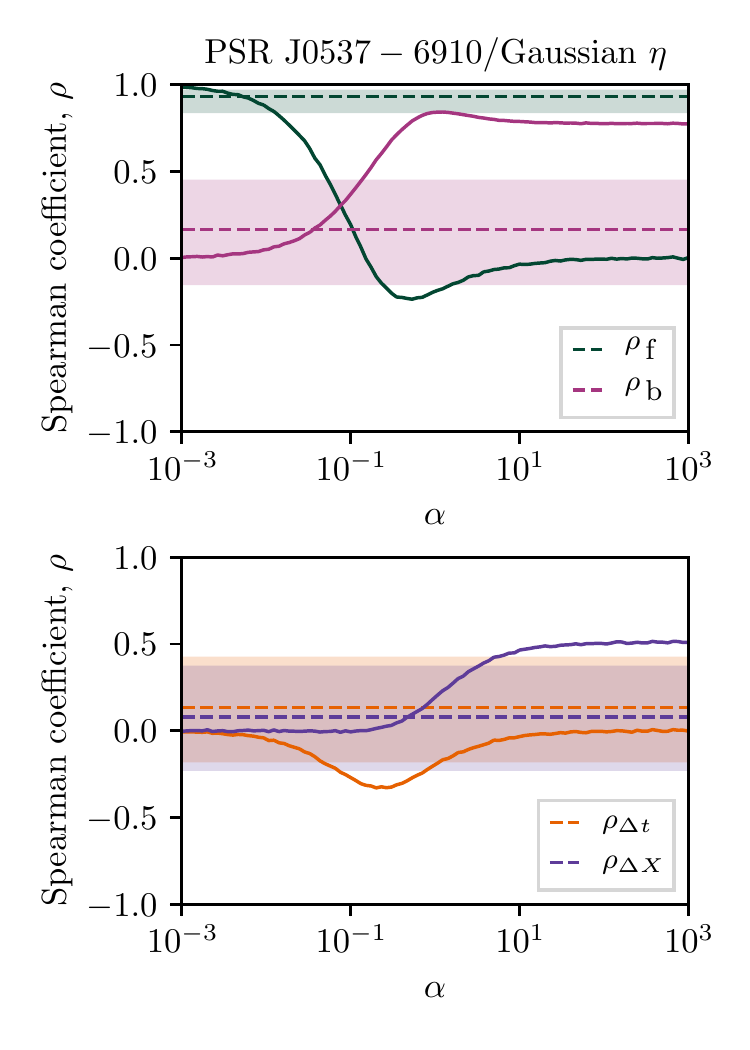}
	\caption{As in Figure \ref{fig:comp_2200}, except for PSR J0537$-$6910, and with a stretchable Gaussian $\eta$ in both panels. Model parameters common to both panels: $\mu_\textrm{G}=0.5$, $\sigma_\textrm{G}=0.125$.}
	\label{fig:comp_6910}
\end{figure}

An example of the above, four-way comparison is presented in Figures \ref{fig:comp_2200} and \ref{fig:comp_6910} for PSR J0534$+$2200 and PSR J0537$-$6910 respectively. The 95\% confidence intervals for the observed correlations are estimated via the standard error for the Spearman correlation coefficient \citep{Bonett2000},
\begin{equation}
{\rho_{\textrm{CI}\pm} \simeq \textrm{tanh}^{-1}\left[\textrm{tanh} \rho \pm 1.96 \left(\frac{1 + \rho^2/2}{\sqrt{N-3}} \right)\right]}\ \ ,
\end{equation}
where $\rho_{\textrm{CI}\pm}$ correspond to the upper and lower limits of the 95\% confidence interval. In Figure \ref{fig:comp_2200}, for PSR J0534$+$2200, the observed forward and backward cross-correlations, as well as the observed waiting time autocorrelation, are consistent with the model, if we have $1 \lesssim \alpha \lesssim 10$ and $\eta$ is a power law. However, the observed negative size autocorrelation contradicts this (and any other) $\alpha$ and $\eta$ combination. On the other hand, in Figure \ref{fig:comp_6910} we see that the observed cross- and autocorrelations for PSR J0537$-$6910 are consistent with the predictions of the model for $\alpha \lesssim 10^{-2}$. The result does not depend on $\eta$; in the low-$\alpha$ regime, all choices of $\eta$ predict the same autocorrelations and cross-correlations; cf. Figure \ref{fig:acorr_sp} and Figure A1 of \citet{Carlin2019b} respectively.

To illustrate roughly what is possible using the method described above, in Table \ref{tab:params} we present ``acceptable'' values of $\alpha$ and functional forms $\eta$ for the five pulsars with the most recorded glitches. We define acceptable values of $\alpha$ to be when, given $\eta$, all four of the forward cross-correlations, backward cross-correlations, size autocorrelations, and waiting time autocorrelations lie within the 95\% confidence interval of the observed correlation coefficients. For PSR J1740$-$3015, when $\eta$ is a power law, any value of $\alpha \gtrsim 1$ is acceptable. For PSR J1341$-$6220, when $\eta$ is a power law, only $0.1 \lesssim \alpha \lesssim 0.5$ is acceptable; however when $\eta$ is a Gaussian, the range shifts to $0.01 \lesssim \alpha \lesssim 0.1$. For PSR J0835$-$4510, when $\eta$ is a power law, any $\alpha \gtrsim 0.5$ is acceptable. When $\eta$ is a Gaussian, $0.05 \lesssim \alpha \lesssim 0.5$ is acceptable. We emphasize that this procedure is not equivalent to a precise parameter estimation or a systematic fit, which we leave to future work given the paucity of current data.

\begin{table}
	\centering %
	\caption{Acceptable values of $\alpha$, given a functional form of $\eta$, for which the state-dependent Poisson process automaton produces cross-correlations and autocorrelations that all lie within the 95\% confidence interval of the correlations observed in five pulsars. See text for details.}
	\begin{tabular}{l l l}
	Name & {Acceptable $\alpha$} & {$\eta$ functional form} \\
	\midrule
	PSR J0534$+$2200  & $1 \lesssim \alpha \lesssim 10$ 	& Power law \\
	PSR J1740$-$3015  & $ \alpha \gtrsim 1$ 				& Power law \\
	PSR J1341$-$6220  & $0.1 \lesssim \alpha \lesssim 0.5$  & Power law \\
		  			  & $0.01 \lesssim \alpha \lesssim 0.1$ & Gaussian \\
	PSR J0537$-$6910  & $\alpha \lesssim 0.01$ 				& Any \\
	PSR J0835$-$4510  & $\alpha \gtrsim 0.5$ 				& Power law \\
		 			  & $0.05 \lesssim \alpha \lesssim 0.5$ & Gaussian \\
	\bottomrule
	\end{tabular}
	\label{tab:params}
\end{table}

As explained by \citet{Melatos2018}, the product of the (observable) long-term average spin-down rate, $\dot{\nu}$, and the average waiting time between glitches for each pulsar, $\langle \Delta t \rangle$, is proportional to $X_\textrm{cr} / \alpha$, if $\eta$ is a separable function of the form
\begin{equation}
	\label{eq:eta_sep}
	{\eta(\Delta X \mid \xp) = (\delta + 1)\left( \xp - \Delta X \right)^\delta \xp^{-(\delta+1)}}\ \ .
\end{equation}
The cross-correlations produced by \eqref{eq:eta_sep} are similar to the standard power law $\eta$, i.e. $\rhof \approx 1$ and $\rhob \ll 1$ for low $\alpha$, $\rhof \ll 1$ and $\rhob \approx 0.15$ for high $\alpha$. We do not use $- \dot{\nu} \langle \Delta t \rangle$ as a proxy for $\alpha^{-1}$ here, because \eqref{eq:eta_sep} produces autocorrelations similar to those of a uniform $\eta$, i.e. $\rhox \ll 1$ and $\rhot \ll 1$ for all $\alpha^{-1}$ and hence all $- \dot{\nu} \langle \Delta t \rangle$.
 
\subsection{Size and waiting time PDFs}
The state-dependent Poisson process models more than just autocorrelations and cross-correlations. For example one can compare the shapes of the waiting time and size PDFs, $p(\Delta t)$ and $p(\Delta X)$, to those observed in real pulsars \citep{Howitt2018, Carlin2019b}. ``Poisson-like'' glitching activity, with exponential $p(\Delta t)$ and power-law $p(\Delta X)$, follows from the state-dependent Poisson process if $\eta$ is a power law, and we have $\alpha \gtrsim 1$. On the other hand, ``quasi-periodic'' glitch activity, with unimodal $p(\Delta t)$ and $p(\Delta X)$ \citep{Melatos2008, Howitt2018}, follows from the state-dependent Poisson process if $\eta$ is unimodal (e.g. Gaussian), and we have $\alpha \lesssim 1$. When $\eta$ is unimodal, and $\alpha \gtrsim 1$, the model generates an exponential $p(\Delta t)$, and a monotonically decreasing $p(\Delta X)$. Whether this adequately describes what is observed for ``Poisson-like'' objects is an open question, given the low ($N\leq24$) number of glitches observed in these objects \citep{Carlin2019b}. 

We now ask whether the added information from autocorrelations in waiting times and sizes helps with input selection and/or parameter estimation when loosely fitting the state-dependent Poisson process to real data as in Section \ref{sec:combine}. The answer is yes, in some cases. For example, modeling all glitching pulsars with a common, unimodal $\eta$ such as a Gaussian is inappropriate, as a unimodal $\eta$ does not generate positive $\rhot$ for any $\alpha$ and only generates positive $\rhox$. Thus the negative $\rhox$ seen in PSR J0534$+$2200 implies that a unimodal $\eta$ cannot adequately model all pulsars. This conclusion cannot be reached from just the shapes of $p(\Delta t)$ and $p(\Delta X)$ \citep{Carlin2019b}, nor the cross-correlations alone \citep{Melatos2018}. In a similar vein, although the cross-correlations, waiting time and size distributions predicted by the state-dependent Poisson process with a power-law $\eta$ are consistent with data from PSR J0534$+$2200, the autocorrelations complicate the picture, because the negative size autocorrelation is not predicted for any value of $\alpha$. 

\subsection{Physical implications}
If the data do not support the same $\eta$ applying to all glitching pulsars, it may mean that different physical mechanisms cause glitches in different pulsars. Different functional forms of $\eta$ are associated with different underlying physical processes. For example, a scale-free power law is characteristic of a spatially correlated knock-on process, such as superfluid vortex avalanches or crustquakes \citep{Middleditch2006, Warszawski2011}. On the other hand, a unimodal function implies a characteristic size for the stress released at each glitch, which is harder to explain microphysically but is consistent with a fluid instability triggered at a critical relative angular velocity \citep{Andersson2003, Mastrano2005, Melatos2007, Glampedakis2009}. Improved measurements of cross-correlations and autocorrelations will improve our ability to discriminate between different functional forms of $\eta$ and therefore different underlying physical mechanisms. Likewise if future glitch observations in PSR J0534$+$2200 continue to show negative size autocorrelations, it becomes hard to reconcile the observations with the state-dependent Poisson process, and the meta-model's applicability to that pulsar should be questioned. In other words, the canonical view that glitches are the result of a marginally critical system, i.e. a process that hovers near an instability threshold, may not be valid in all pulsars.

\section{Conclusions}
As the number of recorded glitches grows it is profitable to disaggregate the data and study the time-ordered nature of the events in individual pulsars. One avenue is to study cross-correlations between the size of a glitch and the waiting time to the next (or previous) glitch \citep{Melatos2018}. Another is to study the autocorrelations between consecutive glitch sizes and waiting times. 

We find no significant autocorrelations between waiting times or sizes in the top five most glitching pulsars, barring perhaps a negative size autocorrelation in PSR J0534$+$2200 ($\rhox = 0.46$, p-value $=0.046$). The absence of autocorrelations is nevertheless informative in the context of the general stress-release meta-model for glitches described by a state-dependent Poisson process \citep{Fulgenzi2017}, which predicts some small autocorrelations under certain conditions. In the fast spin-down regime ($\alpha \lesssim \num{5e-2}$) the meta-model predicts $\rhox \ll 1$ and $\rhot \ll 1$, regardless of the functional form of the conditional jump distribution $\eta(\Delta X \mid \xp)$. If $\eta$ is a power law, as expected for spatially correlated mechanisms such as superfluid vortex avalanches or crustquakes, any nonzero autocorrelations observed become difficult to explain in the context of a state-dependent Poisson process. If on the other hand $\eta$ is unimodal, we expect to see $\rhot < 0$ and $\rhox > 0$ in some pulsars. A unimodal $\eta$ corresponds more closely to a trigger that produces glitches of a characteristic size, e.g. a superfluid instability.

Combining observations of cross-correlations, autocorrelations, and the shapes of the waiting time and size distributions places stronger constraints on the state-dependent Poisson process meta-model than any single statistical measurement. For PSR J0534$+$2200 we find that a power-law $\eta$ and $1 \lesssim \alpha \lesssim 10$ best describe the data, although $\rhox < 0$ complicates the picture. For PSR J0537$-$6910 we find that a Gaussian $\eta$ and $\alpha \lesssim 10^{-2}$ adequately describe the data. For PSR J1740$-$3015 we find a power-law $\eta$ and $\alpha \gtrsim 1$ are acceptable. For PSR1341$-$6220 we find that both a power-law $\eta$ with $0.1 \lesssim \alpha \lesssim 0.5$, and a Gaussian $\eta$ with $0.01 \lesssim \alpha \lesssim 0.1$ adequately describe the data. Finally, for PSR J0835$-$4510 we find that both a power-law $\eta$ with $\alpha \gtrsim 0.5$, and a Gaussian $\eta$ with $0.05 \lesssim \alpha \lesssim 0.5$ are acceptable. 

Precise parameter estimation lies outside the scope of this paper. It involves fitting more than eight independent parameters, along with the functional form of $\eta$, a challenging numerical exercise attempted recently as a proof of principle by \citet{Melatos2019} for three pulsars with $N \geq 23$. Larger data sets are needed for this kind of fitting to become statistically informative. 

\section*{Acknowledgements}
Parts of this research are supported by the Australian Research Council (ARC) Centre
of Excellence for Gravitational Wave Discovery (OzGrav) (project number
CE170100004) and ARC Discovery Project DP170103625. J.B. Carlin is supported by an Australian Postgraduate Award.

%%%%%%%%%%%%%%%%%%%%%%%%%%%%%%%%%%%%%%%%%%%%%%%%%%


\bibliographystyle{mnras}
\bibliography{acorr_bib}

\begin{thebibliography}{}
\makeatletter
\relax
\def\mn@urlcharsother{\let\do\@makeother \do\$\do\&\do\#\do\^\do\_\do\%\do\~}
\def\mn@doi{\begingroup\mn@urlcharsother \@ifnextchar [ {\mn@doi@}
  {\mn@doi@[]}}
\def\mn@doi@[#1]#2{\def\@tempa{#1}\ifx\@tempa\@empty \href
  {http://dx.doi.org/#2} {doi:#2}\else \href {http://dx.doi.org/#2} {#1}\fi
  \endgroup}
\def\mn@eprint#1#2{\mn@eprint@#1:#2::\@nil}
\def\mn@eprint@arXiv#1{\href {http://arxiv.org/abs/#1} {{\tt arXiv:#1}}}
\def\mn@eprint@dblp#1{\href {http://dblp.uni-trier.de/rec/bibtex/#1.xml}
  {dblp:#1}}
\def\mn@eprint@#1:#2:#3:#4\@nil{\def\@tempa {#1}\def\@tempb {#2}\def\@tempc
  {#3}\ifx \@tempc \@empty \let \@tempc \@tempb \let \@tempb \@tempa \fi \ifx
  \@tempb \@empty \def\@tempb {arXiv}\fi \@ifundefined
  {mn@eprint@\@tempb}{\@tempb:\@tempc}{\expandafter \expandafter \csname
  mn@eprint@\@tempb\endcsname \expandafter{\@tempc}}}

\bibitem[\protect\citeauthoryear{{Anderson} \& {Itoh}}{{Anderson} \&
  {Itoh}}{1975}]{Anderson1975}
{Anderson} P.~W.,  {Itoh} N.,  1975, \mn@doi [\nat] {10.1038/256025a0}, \href
  {https://ui.adsabs.harvard.edu/\#abs/1975Natur.256...25A} {256, 25}

\bibitem[\protect\citeauthoryear{{Andersson}, {Comer}  \& {Prix}}{{Andersson}
  et~al.}{2003}]{Andersson2003}
{Andersson} N.,  {Comer} G.~L.,   {Prix} R.,  2003, \mn@doi [\prl]
  {10.1103/PhysRevLett.90.091101}, \href
  {https://ui.adsabs.harvard.edu/abs/2003PhRvL..90i1101A} {90, 091101}

\bibitem[\protect\citeauthoryear{{Antonopoulou}, {Espinoza}, {Kuiper}  \&
  {Andersson}}{{Antonopoulou} et~al.}{2018}]{Antonopoulou2018}
{Antonopoulou} D.,  {Espinoza} C.~M.,  {Kuiper} L.,   {Andersson} N.,  2018,
  \mn@doi [\mnras] {10.1093/mnras/stx2429}, \href
  {https://ui.adsabs.harvard.edu/#abs/2018MNRAS.473.1644A} {473, 1644}

\bibitem[\protect\citeauthoryear{{Arzoumanian} et~al.,}{{Arzoumanian}
  et~al.}{2018}]{Arzoumanian2018}
{Arzoumanian} Z.,  et~al., 2018, \mn@doi [\apjs] {10.3847/1538-4365/aab5b0},
  \href {https://ui.adsabs.harvard.edu/abs/2018ApJS..235...37A} {235, 37}

\bibitem[\protect\citeauthoryear{{Ashton}, {Prix}  \& {Jones}}{{Ashton}
  et~al.}{2017}]{Ashton2017}
{Ashton} G.,  {Prix} R.,   {Jones} D.~I.,  2017, \mn@doi [\prd]
  {10.1103/PhysRevD.96.063004}, \href
  {https://ui.adsabs.harvard.edu/\#abs/2017PhRvD..96f3004A} {96, 063004}

\bibitem[\protect\citeauthoryear{{Bonett} \& {Wright}}{{Bonett} \&
  {Wright}}{2000}]{Bonett2000}
{Bonett} D.~G.,  {Wright} T.~A.,  2000, \mn@doi [Psychometrika]
  {10.1007/BF02294183}, 65, 23

\bibitem[\protect\citeauthoryear{{Buchner} \& {Flanagan}}{{Buchner} \&
  {Flanagan}}{2008}]{Buchner2008}
{Buchner} S.,  {Flanagan} C.,  2008, in {Bassa} C.,  {Wang} Z.,  {Cumming} A.,
   {Kaspi} V.~M.,  eds,  American Institute of Physics Conference Series Vol.
  983, 40 Years of Pulsars: Millisecond Pulsars, Magnetars and More. pp
  145--147, \mn@doi{10.1063/1.2900129}

\bibitem[\protect\citeauthoryear{{Carlin} \& {Melatos}}{{Carlin} \&
  {Melatos}}{2019}]{Carlin2019b}
{Carlin} J.~B.,  {Melatos} A.,  2019, \mn@doi [\mnras] {10.1093/mnras/sty3433},
  \href {https://ui.adsabs.harvard.edu/\#abs/2019MNRAS.483.4742C} {483, 4742}

\bibitem[\protect\citeauthoryear{{Carlin}, {Melatos}  \& {Vukcevic}}{{Carlin}
  et~al.}{2019}]{Carlin2019a}
{Carlin} J.~B.,  {Melatos} A.,   {Vukcevic} D.,  2019, \mn@doi [\mnras]
  {10.1093/mnras/sty2865}, \href
  {https://ui.adsabs.harvard.edu/\#abs/2019MNRAS.482.3736C} {482, 3736}

\bibitem[\protect\citeauthoryear{{Caruso}, {Pluchino}, {Latora}, {Vinciguerra}
  \& {Rapisarda}}{{Caruso} et~al.}{2007}]{Caruso2007}
{Caruso} F.,  {Pluchino} A.,  {Latora} V.,  {Vinciguerra} S.,   {Rapisarda} A.,
   2007, \mn@doi [\pre] {10.1103/PhysRevE.75.055101}, \href
  {http://adsabs.harvard.edu/abs/2007PhRvE..75e5101C} {75, 055101}

\bibitem[\protect\citeauthoryear{Cox}{Cox}{1955}]{Cox1955}
Cox D.~R.,  1955, \mn@doi [J. R. Stat. Soc.] {10.2307/2983950}, 17, 129

\bibitem[\protect\citeauthoryear{{Daly} \& {Porporato}}{{Daly} \&
  {Porporato}}{2007}]{Daly2007}
{Daly} E.,  {Porporato} A.,  2007, \mn@doi [\pre] {10.1103/PhysRevE.75.011119},
  \href {https://ui.adsabs.harvard.edu/\#abs/2007PhRvE..75a1119D} {75, 011119}

\bibitem[\protect\citeauthoryear{{Downs}}{{Downs}}{1981}]{Downs1981}
{Downs} G.~S.,  1981, \mn@doi [\apj] {10.1086/159330}, \href
  {https://ui.adsabs.harvard.edu/abs/1981ApJ...249..687D} {249, 687}

\bibitem[\protect\citeauthoryear{{Espinoza}, {Lyne}, {Stappers}  \&
  {Kramer}}{{Espinoza} et~al.}{2011}]{Espinoza2011}
{Espinoza} C.~M.,  {Lyne} A.~G.,  {Stappers} B.~W.,   {Kramer} M.,  2011,
  \mn@doi [\mnras] {10.1111/j.1365-2966.2011.18503.x}, \href
  {https://ui.adsabs.harvard.edu/\#abs/2011MNRAS.414.1679E} {414, 1679}

\bibitem[\protect\citeauthoryear{{Espinoza}, {Antonopoulou}, {Stappers},
  {Watts}  \& {Lyne}}{{Espinoza} et~al.}{2014}]{Espinoza2014}
{Espinoza} C.~M.,  {Antonopoulou} D.,  {Stappers} B.~W.,  {Watts} A.,   {Lyne}
  A.~G.,  2014, \mn@doi [\mnras] {10.1093/mnras/stu395}, \href
  {https://ui.adsabs.harvard.edu/\#abs/2014MNRAS.440.2755E} {440, 2755}

\bibitem[\protect\citeauthoryear{{Ferdman}, {Archibald}, {Gourgouliatos}  \&
  {Kaspi}}{{Ferdman} et~al.}{2018}]{Ferdman2017}
{Ferdman} R.~D.,  {Archibald} R.~F.,  {Gourgouliatos} K.~N.,   {Kaspi} V.~M.,
  2018, \mn@doi [\apj] {10.3847/1538-4357/aaa198}, \href
  {https://ui.adsabs.harvard.edu/\#abs/2018ApJ...852..123F} {852, 123}

\bibitem[\protect\citeauthoryear{{Fuentes}, {Espinoza}, {Reisenegger}, {Shaw},
  {Stappers}  \& {Lyne}}{{Fuentes} et~al.}{2017}]{Fuentes2017}
{Fuentes} J.~R.,  {Espinoza} C.~M.,  {Reisenegger} A.,  {Shaw} B.,  {Stappers}
  B.~W.,   {Lyne} A.~G.,  2017, \mn@doi [\aap] {10.1051/0004-6361/201731519},
  \href {https://ui.adsabs.harvard.edu/#abs/2017A&A...608A.131F} {608, A131}

\bibitem[\protect\citeauthoryear{{Fulgenzi}, {Melatos}  \& {Hughes}}{{Fulgenzi}
  et~al.}{2017}]{Fulgenzi2017}
{Fulgenzi} W.,  {Melatos} A.,   {Hughes} B.~D.,  2017, \mn@doi [\mnras]
  {10.1093/mnras/stx1353}, \href
  {https://ui.adsabs.harvard.edu/\#abs/2017MNRAS.470.4307F} {470, 4307}

\bibitem[\protect\citeauthoryear{{Glampedakis} \& {Andersson}}{{Glampedakis} \&
  {Andersson}}{2009}]{Glampedakis2009}
{Glampedakis} K.,  {Andersson} N.,  2009, \mn@doi [\prl]
  {10.1103/PhysRevLett.102.141101}, \href
  {https://ui.adsabs.harvard.edu/abs/2009PhRvL.102n1101G} {102, 141101}

\bibitem[\protect\citeauthoryear{Good}{Good}{2006}]{Good2006}
Good P.,  2006, Permutation, Parametric, and Bootstrap Tests of Hypotheses.
Springer Series in Statistics, Springer, New York

\bibitem[\protect\citeauthoryear{{Grandell}}{{Grandell}}{1976}]{Grandell1976}
{Grandell} J.,  1976, Doubly stochastic Poisson processes.
Lecture notes in mathematics: 529, Springer-Verlag, Berlin; New York

\bibitem[\protect\citeauthoryear{Hall}{Hall}{1992}]{Hall1992}
Hall P.,  1992, {The Bootstrap and Edgeworth Expansion}, 1st edn.
Springer-Verlag

\bibitem[\protect\citeauthoryear{{Haskell} \& {Melatos}}{{Haskell} \&
  {Melatos}}{2015}]{Haskell2015}
{Haskell} B.,  {Melatos} A.,  2015, \mn@doi [Int. J. Mod. Phys.]
  {10.1142/S0218271815300086}, \href
  {https://ui.adsabs.harvard.edu/\#abs/2015IJMPD..2430008H} {24, 1530008}

\bibitem[\protect\citeauthoryear{{Howitt}, {Melatos}  \& {Delaigle}}{{Howitt}
  et~al.}{2018}]{Howitt2018}
{Howitt} G.,  {Melatos} A.,   {Delaigle} A.,  2018, \mn@doi [\apj]
  {10.3847/1538-4357/aae20a}, \href
  {https://ui.adsabs.harvard.edu/\#abs/2018ApJ...867...60H} {867, 60}

\bibitem[\protect\citeauthoryear{{Jankowski} et~al.,}{{Jankowski}
  et~al.}{2019}]{Jankowski2019}
{Jankowski} F.,  et~al., 2019, \mn@doi [\mnras] {10.1093/mnras/sty3390}, \href
  {https://ui.adsabs.harvard.edu/abs/2019MNRAS.484.3691J} {484, 3691}

\bibitem[\protect\citeauthoryear{{Janssen} \& {Stappers}}{{Janssen} \&
  {Stappers}}{2006}]{Janssen2006}
{Janssen} G.~H.,  {Stappers} B.~W.,  2006, \mn@doi [\aap]
  {10.1051/0004-6361:20065267}, \href
  {https://ui.adsabs.harvard.edu/\#abs/2006A&A...457..611J} {457, 611}

\bibitem[\protect\citeauthoryear{{Larson} \& {Link}}{{Larson} \&
  {Link}}{2002}]{Larson2002}
{Larson} M.~B.,  {Link} B.,  2002, \mn@doi [\mnras]
  {10.1046/j.1365-8711.2002.05439.x}, \href
  {https://ui.adsabs.harvard.edu/\#abs/2002MNRAS.333..613L} {333, 613}

\bibitem[\protect\citeauthoryear{Lehmann \& D'Abrera}{Lehmann \&
  D'Abrera}{2006}]{Lehmann2006}
Lehmann E.,  D'Abrera H.,  2006, Nonparametrics: Statistical Methods Based on
  Ranks.
Springer, New York

\bibitem[\protect\citeauthoryear{{Lyne}, {Shemar}  \& {Smith}}{{Lyne}
  et~al.}{2000}]{Lyne2000}
{Lyne} A.~G.,  {Shemar} S.~L.,   {Smith} F.~G.,  2000, \mn@doi [\mnras]
  {10.1046/j.1365-8711.2000.03415.x}, \href
  {http://adsabs.harvard.edu/abs/2000MNRAS.315..534L} {315, 534}

\bibitem[\protect\citeauthoryear{{Lyne}, {Jordan}, {Graham-Smith}, {Espinoza},
  {Stappers}  \& {Weltevrede}}{{Lyne} et~al.}{2015}]{Lyne2015}
{Lyne} A.~G.,  {Jordan} C.~A.,  {Graham-Smith} F.,  {Espinoza} C.~M.,
  {Stappers} B.~W.,   {Weltevrede} P.,  2015, \mn@doi [\mnras]
  {10.1093/mnras/stu2118}, \href
  {https://ui.adsabs.harvard.edu/\#abs/2015MNRAS.446..857L} {446, 857}

\bibitem[\protect\citeauthoryear{{Mastrano} \& {Melatos}}{{Mastrano} \&
  {Melatos}}{2005}]{Mastrano2005}
{Mastrano} A.,  {Melatos} A.,  2005, \mn@doi [\mnras]
  {10.1111/j.1365-2966.2005.09219.x}, \href
  {https://ui.adsabs.harvard.edu/abs/2005MNRAS.361..927M} {361, 927}

\bibitem[\protect\citeauthoryear{{Melatos} \& {Drummond}}{{Melatos} \&
  {Drummond}}{2019}]{Melatos2019}
{Melatos} A.,  {Drummond} L.~V.,  2019, submitted to ApJ for publication

\bibitem[\protect\citeauthoryear{{Melatos} \& {Peralta}}{{Melatos} \&
  {Peralta}}{2007}]{Melatos2007}
{Melatos} A.,  {Peralta} C.,  2007, \mn@doi [\apj] {10.1086/518598}, \href
  {https://ui.adsabs.harvard.edu/abs/2007ApJ...662L..99M} {662, L99}

\bibitem[\protect\citeauthoryear{{Melatos}, {Peralta}  \& {Wyithe}}{{Melatos}
  et~al.}{2008}]{Melatos2008}
{Melatos} A.,  {Peralta} C.,   {Wyithe} J.~S.~B.,  2008, \mn@doi [\apj]
  {10.1086/523349}, \href
  {https://ui.adsabs.harvard.edu/\#abs/2008ApJ...672.1103M} {672, 1103}

\bibitem[\protect\citeauthoryear{{Melatos}, {Howitt}  \& {Fulgenzi}}{{Melatos}
  et~al.}{2018}]{Melatos2018}
{Melatos} A.,  {Howitt} G.,   {Fulgenzi} W.,  2018, \mn@doi [\apj]
  {10.3847/1538-4357/aad228}, \href
  {https://ui.adsabs.harvard.edu/\#abs/2018ApJ...863..196M} {863, 196}

\bibitem[\protect\citeauthoryear{{Middleditch}, {Marshall}, {Wang}, {Gotthelf}
  \& {Zhang}}{{Middleditch} et~al.}{2006}]{Middleditch2006}
{Middleditch} J.,  {Marshall} F.~E.,  {Wang} Q.~D.,  {Gotthelf} E.~V.,
  {Zhang} W.,  2006, \mn@doi [\apj] {10.1086/508736}, \href
  {https://ui.adsabs.harvard.edu/\#abs/2006ApJ...652.1531M} {652, 1531}

\bibitem[\protect\citeauthoryear{{Ng}}{{Ng}}{2018}]{Ng2018}
{Ng} C.,  2018, in {Weltevrede} P.,  {Perera} B.~B.~P.,  {Preston} L.~L.,
  {Sanidas} S.,  eds,  IAU Symposium Vol. 337, Pulsar Astrophysics the Next
  Fifty Years. pp 179--182 (\mn@eprint {arXiv} {1711.02104}),
  \mn@doi{10.1017/S1743921317010638}

\bibitem[\protect\citeauthoryear{{Paczuski}, {Boettcher}  \&
  {Baiesi}}{{Paczuski} et~al.}{2005}]{Paczuski2005}
{Paczuski} M.,  {Boettcher} S.,   {Baiesi} M.,  2005, \mn@doi [Physical Review
  Letters] {10.1103/PhysRevLett.95.181102}, \href
  {http://adsabs.harvard.edu/abs/2005PhRvL..95r1102P} {95, 181102}

\bibitem[\protect\citeauthoryear{{Palfreyman}, {Dickey}, {Ellingsen}, {Jones}
  \& {Hotan}}{{Palfreyman} et~al.}{2016}]{Palfreyman2016}
{Palfreyman} J.~L.,  {Dickey} J.~M.,  {Ellingsen} S.~P.,  {Jones} I.~R.,
  {Hotan} A.~W.,  2016, \mn@doi [\apj] {10.3847/0004-637X/820/1/64}, \href
  {https://ui.adsabs.harvard.edu/abs/2016ApJ...820...64P} {820, 64}

\bibitem[\protect\citeauthoryear{{Santra}, {Chanu}  \& {Deb}}{{Santra}
  et~al.}{2007}]{Santra2007}
{Santra} S.~B.,  {Chanu} S.~R.,   {Deb} D.,  2007, \mn@doi [\pre]
  {10.1103/PhysRevE.75.041122}, \href
  {http://adsabs.harvard.edu/abs/2007PhRvE..75d1122S} {75, 041122}

\bibitem[\protect\citeauthoryear{{Shemar} \& {Lyne}}{{Shemar} \&
  {Lyne}}{1996}]{Shemar1996}
{Shemar} S.~L.,  {Lyne} A.~G.,  1996, \mn@doi [\mnras]
  {10.1093/mnras/282.2.677}, \href
  {http://adsabs.harvard.edu/abs/1996MNRAS.282..677S} {282, 677}

\bibitem[\protect\citeauthoryear{{Utsu}, {Ogata}, {S}  \& {Matsu'ura}}{{Utsu}
  et~al.}{1995}]{Utsu1995}
{Utsu} T.,  {Ogata} Y.,  {S} R.,   {Matsu'ura} 1995, \mn@doi [J. Phys. Earth]
  {10.4294/jpe1952.43.1}, 43, 1

\bibitem[\protect\citeauthoryear{{Warszawski} \& {Melatos}}{{Warszawski} \&
  {Melatos}}{2011}]{Warszawski2011}
{Warszawski} L.,  {Melatos} A.,  2011, \mn@doi [\mnras]
  {10.1111/j.1365-2966.2011.18803.x}, \href
  {https://ui.adsabs.harvard.edu/\#abs/2011MNRAS.415.1611W} {415, 1611}

\bibitem[\protect\citeauthoryear{{Wheatland}}{{Wheatland}}{2008}]{Wheatland2008}
{Wheatland} M.~S.,  2008, \mn@doi [\apj] {10.1086/587871}, \href
  {https://ui.adsabs.harvard.edu/\#abs/2008ApJ...679.1621W} {679, 1621}

\bibitem[\protect\citeauthoryear{{Yu} \& {Liu}}{{Yu} \& {Liu}}{2017}]{Yu2017}
{Yu} M.,  {Liu} Q.~J.,  2017, \mn@doi [\mnras] {10.1093/mnras/stx702}, \href
  {https://ui.adsabs.harvard.edu/\#abs/2017MNRAS.468.3031Y} {468, 3031}

\bibitem[\protect\citeauthoryear{{Yu} et~al.,}{{Yu} et~al.}{2013}]{Yu2013}
{Yu} M.,  et~al., 2013, \mn@doi [\mnras] {10.1093/mnras/sts366}, \href
  {https://ui.adsabs.harvard.edu/\#abs/2013MNRAS.429..688Y} {429, 688}

\bibitem[\protect\citeauthoryear{{de Menech} \& {Stella}}{{de Menech} \&
  {Stella}}{2000}]{DeMenech2000}
{de Menech} M.,  {Stella} A.~L.,  2000, \mn@doi [\pre]
  {10.1103/PhysRevE.62.R4528}, \href
  {https://ui.adsabs.harvard.edu/abs/2000PhRvE..62.4528D} {62, R4528}

\makeatother
\end{thebibliography}

%%%%%%%%%%%%%%%%%%%%%%%%%%%%%%%%%%%%%%%%%%%%%%%%%%

%%%%%%%%%%%%%%%%% APPENDICES %%%%%%%%%%%%%%%%%%%%%

% \appendix
%%%%%%%%%%%%%%%%%%%%%%%%%%%%%%%%%%%%%%%%%%%%%%%%%%

% Don't change these lines
\bsp	% typesetting comment
\label{lastpage}
\end{document}